\documentclass[sigconf]{ acmart}
\setcopyright{none}
\acmYear{2023}
\usepackage{amsmath,amsfonts}
\usepackage{algorithmic}
\usepackage{graphicx}
\usepackage{textcomp}
\usepackage{xcolor}
\newcommand{\tool}{\textsc{GraphCodeAttack}}
\usepackage{tikz}
\usepackage{bbm}
\usepackage[ruled]{algorithm2e}
\usepackage{tcolorbox}
\usepackage{multirow}
\usepackage{booktabs}
\usepackage{url}
\usepackage{caption}
\usepackage{hyperref}
\usepackage{balance}
\usepackage{multicol}

\usetikzlibrary{positioning, shapes.geometric, arrows.meta, calc}
\newtcolorbox{graybox}{
  colback=gray!20,
  colframe=black,
  boxrule=2pt, % Adjust this value for border thickness
  arc=5mm,     % Adjust this value for rounded corners radius
  boxsep=2pt
}

\author{Thanh-Dat Nguyen}
\affiliation{
  \institution{University of Melbourne}
  \city{Melbourne}
  \country{Australia}
}
\email{thanhdatn@student.unimelb.edu.au}

\author{Yang Zhou}
\affiliation{
  \institution{Singapore Management University}
  \city{Singapore}
  \country{Singapore}
}
\email{zyang@smu.edu.sg}

\author{Xuan-Bach D. Le}
\affiliation{
  \institution{University of Melbourne}
  \city{Melbourne}
  \country{Australia}
}
\email{back.le@unimelb.edu.au}

\author{Patanamon (Pick) Thongtanunam}
\affiliation{
  \institution{University of Melbourne}
  \city{Melbourne}
  \country{Australia}
}
\email{patanamon.t@unimelb.edu.au}

\author{David Lo}
\affiliation{
  \institution{Singapore Management University}
  \city{Singapore}
  \country{Singapore}
}
\email{davidlo@smu.edu.sg}

\begin{document}

\title{Adversarial Attacks on Code Models with Discriminative Graph Patterns}

\begin{abstract}

Pre-trained language models of code are now widely used in various software engineering tasks such as code generation, code completion, vulnerability detection, etc. 
This, in turn, poses security and reliability risks to these models.
One of the important threats is \textit{adversarial attacks}, which can lead to erroneous predictions and largely affect model performance on downstream tasks, necessitating a thorough study of adversarial robustness on code models.
Current adversarial attacks on code models usually adopt fixed sets of program transformations, such as variable renaming and dead code insertion. 
Additionally, expert efforts are required to handcraft these transformations and are limited in terms of creating more complex semantic-preserving transformations.

To address the aforementioned challenges, we propose a novel adversarial attack framework, \tool{}, to better evaluate the robustness of code models. Given a target code model, \tool{} automatically mines important code patterns, which can influence the model's decisions, to perturb the structure of input code to the model. To do so, \tool{} uses a set of input source codes to probe the model's outputs. From these source codes and outputs, \tool{} identifies the \textit{discriminative} ASTs patterns that can influence the model decisions. \tool{} then selects appropriate AST patterns, concretizes the selected patterns as attacks, and inserts them as dead code into the model's input program. To effectively synthesize attacks from AST patterns, \tool{} uses a separate pre-trained code model to fill in the ASTs with concrete code snippets.
We evaluate the robustness of two popular code models (e.g., CodeBERT and GraphCodeBERT) against our proposed approach on three tasks: Authorship Attribution, Vulnerability Prediction, and Clone Detection. The experimental results suggest that our proposed approach significantly outperforms state-of-the-art approaches in attacking code models such as CARROT and ALERT. Based on the average attack success rate (ASR), \tool{} achieved $30\%$ improvement over CARROT and $33\%$ improvement over ALERT respectively. Notably, in terms of ASR on GraphCodeBERT model and on Authorship Attribution, \tool{} achieved an ASR of $0.841$, significantly outperforming CARROT and ALERT (with ASR of $0.598$ and $0.615$ respectively).  
\end{abstract}


\maketitle

\section{Introduction}
Code models~\cite{bert, gpt-2, keskar2019ctrl}, especially those built on advanced deep learning architectures, have become increasingly popular recently due to their ability to effectively comprehend programming languages by learning from large-scale code data~\cite{feng2020codebert, guo2021graphcodebert, wang2021codet5}. 
These models have been employed and demonstrated strong performance in various applications, including code completion \cite{liu2020multitask}, 
vulnerability detection~\cite{zhou2019devign}, authorship attribution~\cite{alasumi2017lstmauthorship}, and code clone detection \cite{lu2021codexglue}. 
Despite their success, recent studies have shown that code models are not robust to \textit{adversarial perturbations} \cite{zhang2022Carrot, Zhou2022ALERT} -- i.e., semantic-preserving transformations (e.g., renaming the variables or adding some dead code) of the input, that make a code model change predictions from correct to wrong.

The vulnerability of code models to adversarial perturbations can have serious implications for the security and reliability of downstream tasks that employ these models. Consider a scenario where a code model is integrated into an open-source library's contribution review process to detect and prevent the inclusion of vulnerable or malicious code. In this situation, ill-intentioned actors could craft adversarial perturbations to exploit the weaknesses of the code model, thereby causing it to falsely accept their malicious contributions~\cite{ladisa2022taxonomy}.
As a consequence, the library could unknowingly incorporate security vulnerabilities or harmful code, leading to significant risks for the library's users and potentially damaging the reputation of the project maintainers. This threat model highlights the importance of assessing and enhancing the robustness of code models against adversarial attacks.

In this paper, we evaluate the code model robustness in a black-box setting: the attacker only has access to the model's output and cannot access the internal information (e.g., parameter and gradient information) of the victim model, nor the ground truth label. 
The black-box assumption is realistic as such code models are usually deployed remotely and can be accessed by APIs.
Many recent works~\cite{Zhou2022ALERT, zhang2022Carrot, Yefet2020, jha2023codeattack} also adopt the same assumption.
ALERT~\cite{Zhou2022ALERT} and CARROT~\cite{zhang2022Carrot} are state-of-the-art techniques for adversarial attacks on code models. 
They focus on using a fixed set of hand-crafted patterns to transform the inputs of code models. ALERT~\cite{Zhou2022ALERT} uses variable renaming and CARROT~\cite{zhang2022Carrot} uses additional transformations, e.g., adding dead-code with manually-designed patterns such as $\texttt{while(false)}$, $\texttt{if}\texttt{(false)}$, etc. 
Attacking code models using hand-crated transformations, however, presents certain limitations. Particularly, the hand-crafted patterns may not stay abreast of fastly growing datasets to adequately represent the diverse range of real-world code structures and may have limitations in modeling complex semantic-preserving transformations.
Hence, there is a need for an automated and systematic process of identifying potential adversarial attacks, which will enable the developers thoroughly test and assure the reliability of the code model before releasing it to the users.

%there is no guarantee that they adequately represent the diverse range of real-world code structures. In other words, the fixed sets of transformations may not stay abreast of frequently growing datasets. 

%which can impact their effectiveness. Moreover, the inflexibility of these methods can result in reduced adaptability, limiting their ability to reveal potential model vulnerabilities and defend against new, domain-specific attacks launched by adversaries.

We introduce \tool{}, a novel approach to attack models of code by using \emph{automatically mined} code patterns that can highly influence a target model's decisions. Doing so allows \tool{} to flexibly adapt to different code models with varying training data, as opposed to the use of handcrafted transformations by current state-of-the-art approaches. Given a target model of code and a set of probing data (i.e., a set of data used to test the model's output), \tool{} works in three phases: mining \emph{highly influential} patterns, synthesizing attacks from patterns, and selecting appropriate attacks. 

\tool{} first automatically identifies \emph{discriminative} AST patterns from the probing data (i.e., programs' source code) that highly influence the target model's prediction. To achieve this, we employ a discriminative subgraph mining technique, namely the gspan-CORK algorithm \cite{Thoma2010}. This allows us to find frequent subgraphs or patterns in the data that are discriminative between different classes or groups of data. Second, \tool{} synthesizes concrete attacks based on the patterns mined in the previous step. Note that the mined AST patterns primarily contain structural information, such as node types and edge types, without any actual concrete content (e.g., specific identifiers or particular binary operations among expressions like \texttt{+, -}, etc.). To fill this gap, we leverage a language model, which is different from the model under attack, to synthesize concrete code from abstract patterns. Large language models, such as CodeBERT~\cite{feng2020codebert} and CodeT5~\cite{bui2022detectlocalizerepair}, have demonstrated capabilities in completing code spans that are contextually coherent. We thus leverage these models for the attack synthesis step. \tool{} then searches through the synthesized concrete attacks to find appropriate attacks to be inserted into a given program as input to the target model.

We evaluate the robustness of two popular code models (e.g., CodeBERT and GraphCodeBERT) against our proposed approach on three tasks: Authorship Attribution, Vulnerability Prediction, and Clone Detection. Experiments suggest that our proposed approach significantly outperforms state-of-the-art approaches in code model attacks such as CARROT and ALERT. Based on the average attack success rate (ASR), \tool{} achieved $30\%$ improvement over CARROT and $33\%$ improvement over ALERT respectively. Notably, on the GraphCodeBERT model and on Authorship Attribution, \tool{} achieved a score of $0.84$ in terms of average ASR.  

In summary, we present \tool{}, a novel approach for attacking models of code by synthesizing adversarial examples from attack patterns that are automatically mined from a set of probing data and the corresponding model's output. Our contributions can be summarized as follows:
\begin{itemize}
    \item Our novel approach \tool{} automatically mines attack patterns from a model and the set of probing data, rendering the derived attacks flexibly adaptable to specific target models and domains.
    \item We introduce a novel method that leverages pre-trained language models to automatically generate effective concrete attacks from discovered abstract AST patterns. In comparison with CARROT's random identifier renaming and ALERT's code-model-based identifier renaming, \tool{} surpasses their performance in 5 out of the 6 task-and-model combinations evaluated.
    \item We demonstrate the effectiveness of our approach through extensive experiments, showing that \tool{} can successfully synthesize adversarial examples that challenge the robustness and reliability of code models. Particularly, \tool{} achieved $30\%$ improvement over CARROT and $33\%$ improvement over ALERT on average ASR measurement. Notably, on the GraphCodeBERT model, \tool{} achieved $0.84$ and $0.799$ in terms of ASR on Authorship Attribution and Vulnerability Prediction respectively.  
\end{itemize}

The rest of the paper is organized as follows: Section~\ref{sec:background}  provides background on code models, adversarial attacks, and abstract syntax trees. Section~\ref{sec:methodology} details the proposed \tool{} methodology, including the process of mining AST patterns, and the usage of pre-trained language models for pattern insertion. Section~\ref{sec:rel_work} presents the related works on the problem of adversarial attack on the model of code. Sections~\ref{sec:exp_settings} and~\ref{sec:rqs} describe our experiment settings and the results respectively.

\section{Background}
\label{sec:background}
In this section, we provide essential background information on code models and the use of Abstract Syntax Trees (ASTs) for graph mining techniques that form the basis of our proposed system architecture for attacking code models.

\subsection{Code Models}

Code models \cite{feng2020codebert, guo2021graphcodebert} are machine learning models designed to analyze, understand, and generate source code. They play a crucial role in various software engineering tasks, such as code completion, code summarization, bug detection, and vulnerability identification. Recent advances in deep learning have led to the development of more sophisticated code models, such as Transformer-based models, that can capture complex patterns and structures in source code. These transformer models can be pre-trained using unlabelled code datasets to capture the semantic relations in the source code~\cite{feng2020codebert, guo2021graphcodebert}. After pre-training, these models can be fine-tuned to achieve state-of-the-art performance on downstream tasks such as code completion, vulnerability prediction, authorship attribution, etc.~\cite{lu2021codexglue, codex}.  However, these models are also susceptible to adversarial attacks, where carefully crafted perturbations in the input source code can cause them to produce incorrect predictions or outputs~\cite{Zhou2022ALERT, zhang2022Carrot}. 

\subsection{Graph Mining via Abstract Syntax Trees}

\noindent \textbf{Abstract Syntax Tree}. Abstract Syntax Trees (ASTs) represent the syntactic structure of source code, with nodes corresponding to language constructs and edges indicating the relationships between nodes.
In our context, we describe the ASTs as graphs. 
In detail, an AST is denoted as a graph $G = (V, E)$, where the vertex set $V$ consists of nodes corresponding to language constructs, and the edge set $E$ includes directed edges indicating their parent-child relationships. 
Each node $n \in V$ is associated with a label $l_n$ representing the language construct it denotes, such as variables, expressions, or control structures. 
Similarly, each edge $e \in E$ can also have a corresponding label $l_e$ specifying the relationship between the connected nodes, such as data or control dependencies. 
For example, in an AST representing a simple \texttt{if-else} statement, the nodes might represent the \texttt{if} keyword, the condition, and the branches, while the edges indicate the parent-child relationships among these nodes.
The structure of the AST captures the hierarchical organization and the syntactic dependencies in the source code, which are essential for mining discriminative patterns.

\noindent \textbf{Discriminative Patterns Mining}. 
Discriminative subgraph mining~\cite{Thoma2010} is a branch of graph mining that focuses on discovering subgraphs or patterns that exhibit significant differences between classes in the dataset. 
The goal is to identify the most distinguishing substructures for each class, which can then be used for tasks such as classification, clustering, and anomaly detection. 
Since these subgraphs are discriminative, their presence might be more likely to change the prediction of the target model. Thus, \tool{} perturb the input source code by inserting these patterns.

\tool{} works by finding the  most discriminative subgraphs from a dataset and uses these subgraphs to structurally perturb the input code. Our \tool{} analyzes ASTs since ASTs provide a more structured and semantically rich representation of source code than raw text, facilitating pattern mining and analysis.
Also, ASTs are a generic representation, allowing our approach to be applicable across different programming languages.

\section{Methodology} \label{sec:methodology}

In this section, we present the architecture of \tool{}. \tool{} aims at deriving adversarial attacks of models of code by automatically mining attack patterns from a set of probing data and the corresponding model's output. The attack patterns are in the form of abstract syntax trees (ASTs) that can be used to perturb the structure of a model's input code which influence the model's decisions.

\tool{} has to overcome three primary challenges: (1) How to automatically obtain the set of effective abstract patterns that are flexibly adaptable to each model using a set of probing data (2) How to effectively derive concrete attacks from an abstract pattern, and (3) Which patterns to select and where to insert it into the input code.

To address these challenges, \tool{} operates in three main phases. First, \tool{} formulates the problem of identifying effective AST patterns as \emph{discriminative subgraph mining} (the mining phase~\ref{sec:mining}). Taking a set of input code $\mathcal{D}$ and the corresponding model's predictions as input, \tool{} identifies a set of discriminative abstract syntax tree patterns $\mathcal{P}_A$ that are highly correlated with the model's predictions. 
This means that the presence or absence of these patterns is closely linked to specific predictions made by the models based on the preprocessing phase.
Based upon this link, \tool{} tries to alter the model prediction by inserting these patterns into the input source code. Note that these patterns are inserted to a program in a way that the semantics of the underlying program remain intact. 
% \bach{What do you mean by discriminative graphs towards model prediction? In terms of English, I don't understand}. The combination of these graphs can likely influence the output of the model \bach{Why can they influence the model's out?}. 

Recall that the mined AST patterns are abstract, only containing information such as node type, edge types, etc, without concrete content (e.g., specific identifier, specific binary, unary operations like \texttt{+, -, >, <}, etc.). To synthesize concrete attacks from the abstract AST patterns, \tool{} convert each pattern into a textual form, in which parts that need to be filled in are indicated by a special token \texttt{<MASK>}. This textual representation help facilitate the insertion of the pattern in the attack phase.

%This is done by identifying changeable nodes in the patterns and the unchanged parts from the dataset. The changed parts are denoted with a special token \texttt{<MASK>} and the unchanged parts are kept with the original tokens.

%\dat{High level idea: Finally, given these ast patterns and their corresponding textual patterns. The attack phases aims at }

Finally, the attack phase focuses on determining the valid perturbation of the input source code that makes the target model change prediction.
To search for this perturbation, \tool{} formulates the problem as a search problem of positions in the source code and the corresponding modifications to perform at each corresponding position. 
We sample the positions based on a calculated important score, which specifies the effectiveness of having the statement and not having the statement on the target model's prediction. Followed by that, we estimate the most impactful pattern to insert based on a meta-model over the model's output. Having determined the position as well as the corresponding patterns, we insert these patterns into the input source code.

To perform the attack, we implement a greedy strategy. At each greedy step, we choose the position/pattern combination that can most reduce the confidence of the target model on its output. This assumption has also been adopted by earlier works~\cite{Zhou2022ALERT, zhang2022Carrot}.
We keep track of the number of model queries and stop this process once the maximum number of queries is reached.

We explain in detail the three main phases of \tool{} in the below subsections.

\begin{figure}[t]
    \centering
    \includegraphics[width=\linewidth]{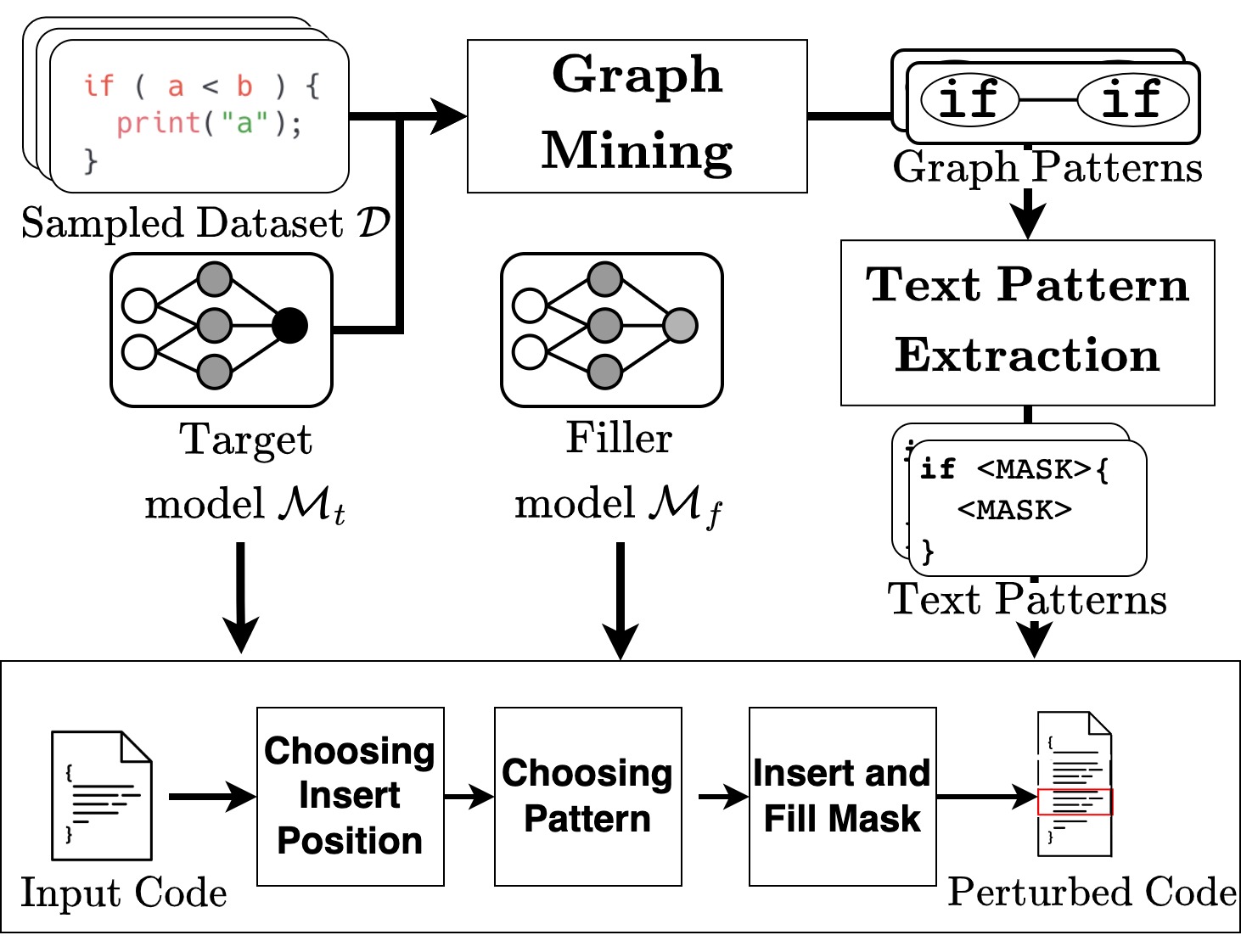}
    \caption{Overview of \tool{}'s method. $\mathcal{M}_t$ is the target victim model, $\mathcal{M}_f$ is the language model used to fill in the \texttt{<MASK>}}
\end{figure}
\begin{figure*}
    \centering
    \includegraphics[width=1.0\textwidth, scale=1.0]{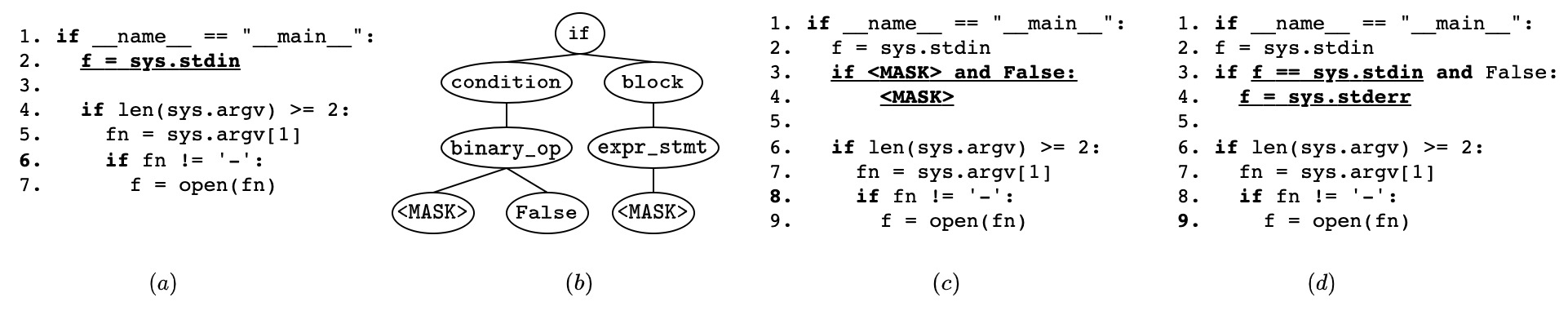}
    \caption{Attacking with pattern: Given the original source code (a), \tool{} identify the important statement on line 2: $\texttt{f = sys.stdin}$. \tool{} then chooses the pattern $(b)$ consisting of an if statement with unknown condition and body. \tool{} inserts this text pattern in the code, resulting in the masked code $(c)$. Finally, \tool{} uses the filler language model $\mathcal{M}_f$ to fill in the mask in $(c)$, resulting in the perturbed code $(d)$ that changes model prediciton}
    \label{fig:motiv_example}
\end{figure*}

\subsection{Mining Attack Patterns}
\label{sec:mining}

In this section, we detail the process of mining attack patterns in \tool{}. The primary objective is to identify a set of discriminative subgraphs $\mathcal{P}_A$ that can effectively influence the target model's prediction.

\subsubsection{Discriminative Subgraph Mining}
\label{sec:discriminative_subgraph_mining}

Given a set of input code samples $S = {s_1, s_2, \dots, s_n}$ and their corresponding model predictions $Y = {y_1, y_2, \dots, y_n}$, we formulate the task of finding effective patterns to attack the model as a discriminative subgraphs mining problem. The goal is to discover a set of subgraphs $\mathcal{P}_A = {P_1, P_2, \dots, P_k}$ that are significantly discriminative. These subgraphs, which discriminate between different classes or groups of data, empower us to effectively sway the model's prediction.

To achieve this, we first construct each AST representation $T_i$ for each code sample $s_i \in S$.  
After obtaining the resulting ASTs set $\mathcal{T} = \{T_1, T_2, \ldots, T_n\}$, we apply the gSpan-CORK algorithm to find the set of subgraphs $\mathcal{P}_A$ that exhibit a high discriminative power. 
gSpan-CORK aims to greedily find the the set of subgraphs $\mathcal{P}_A = \{P_{A_1}, P_{A_2}, \ldots, P_{A_k}\}$ that maximizes the quality criterion $q$. For two classes $0$ and $1$ with the corresponding set of ASTs $\mathcal{T}_{0}$ and $\mathcal{T}_{1}$:
\begin{equation}
    q(\mathcal{P}_A) = - \sum_{P \in \mathcal{P}_A} ( |\mathcal{T}_{0, \bar{P}}| \cdot |\mathcal{T}_{1, \bar{P}}| + |\mathcal{T}_{0, P}| \cdot |\mathcal{T}_{1, P}|)
\end{equation}
where $\mathcal{T}_{0, \bar{P}}$, $\mathcal{T}_{1, \bar{P}}$ are the sets of ASTs belonging to class $0$ and class $1$ that do not contain subgraph $P$. $\mathcal{T}_{0, P}$, $\mathcal{T}_{1, P}$ is the set of ASTs belonging to class $0$ and class $1$ that contains $P$ respectively.
The intuition behind this function is to maximize the difference in subgraph patterns between the two classes.

The first term counts the cases where subgraph $P$ is absent in both class $0$ and class $1$, while the second term counts the cases where subgraph $P$ is present in both classes. By minimizing the sum of these cross-interactions, we aim to find the set of subgraphs $\mathcal{P}_A$ that best differentiates the two classes.
These subgraphs provide the basis for the subsequent attack phase, where they are used to perturb the source code and alter the model's output.
We note that there can be cases where the model can perform multi-class classifications that predict the output to be one of $C$ classes. In this case, we simply construct $C$ one-versus-all graph datasets.

\subsection{Synthesizing Concrete Attacks from AST Patterns}
Recall that each pattern $P$ in the discriminative AST pattern set $\mathcal{P}_A$ is a subgraph. Utilizing these patterns to guide the perturbation of code presents some challenges, as the process is not straightforward. Particularly, as the mined graphs are abstract, we need to synthesize concrete code snippets from the patterns in a way that the snippets are contextually coherent with the underlying program.

\begin{figure}
    \centering
    \includegraphics[width=\linewidth]{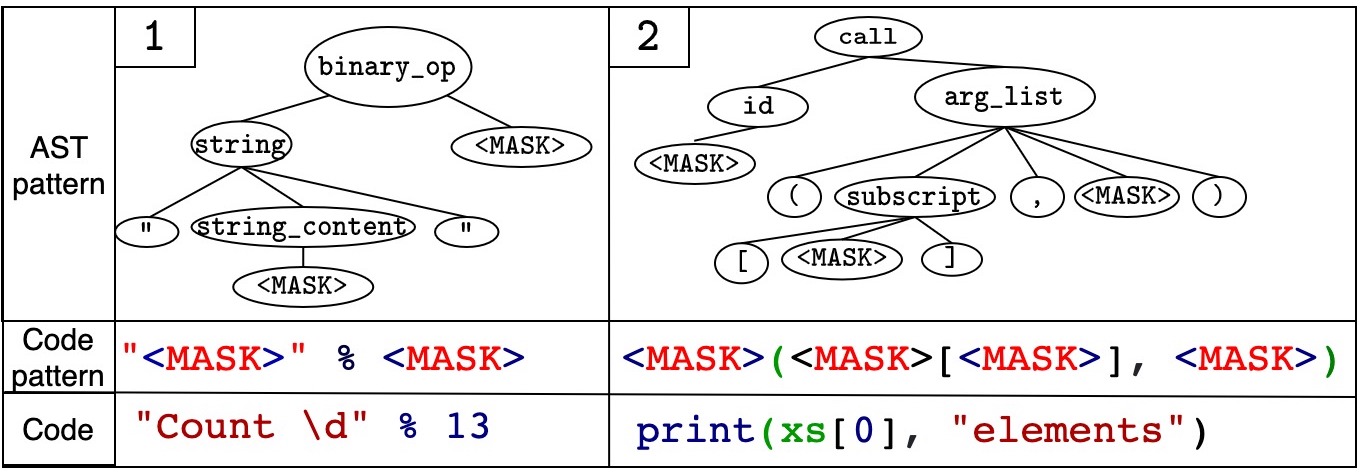}
    \caption{Example of corresponding AST pattern and textual pattern}
    \label{fig:ast_hole}
\end{figure}

Figure~\ref{fig:ast_hole} demonstrates a motivating example. We have an attack pattern consisting of a binary operation with the left side component pointing to a string of unknown content and the right side component pointing to an unknown node. There are several problems with inserting this pattern into the code:
(1) Filling in the content of the right node is problematic as we do not know the actual type, name, or value of the node.
(2) We do not know what exactly the operation of \textbf{Binary Op} node and (3) 
Furthermore, we also have to consider what variable or expression is based on each context to be put in the right side.
\tool{} tackles these challenges using a pre-trained model of code. As an example, consider the abstract syntax subtree depicted in Figure~\ref{fig:ast_hole}. 

To determine which operations can be put in \textbf{Binary Op} node, we identify the corresponding instance of the pattern in the actual source code and narrow down the set of values that can be put in. 
Having identified the instances of the pattern in the dataset, we also know how to identify the components that can be changed. 
For incomplete components (e.g., string without string content, a right node of the binary component), we identify the textual span of the corresponding component and replace its textual content with the special $\texttt{<MASK>}$ token.

The result is the textual representation of the mined subgraph with unknown components replaced by the $\texttt{<MASK>}$ token. As we will see later, this representation facilitates the insertion of the pattern into the source code using the pre-trained language models.

\subsection{Attacking with mined patterns}
Having obtained the pattern and the corresponding textual representation, we proceed to perform the adversarial attack on the target model.
Taking a source code as an input, \tool{} repeatedly chooses the most important statements along with a pattern that likely impacts the model prediction and inserts the pattern next to the statement until it reaches the token threshold limit. We give an illustration in Figure ~\ref{fig:motiv_example}.

\subsubsection{Statement-level important score estimation} 
To choose the most important statement, we quantify the impact of a statement by computing the difference in the target model's probability estimates with and without the inclusion of the statement. Let $P_{\mathcal{M}_t, c_t}(s)$ denote the target model $\mathcal{M}_t$'s output probability for a given input code sample $s$ and for the target class $c_t$. 
Let $a$ be a statement in $s$, the impact of $a$ on $s$ is:
\begin{equation}
    \Delta P_{\mathcal{M}_t, c_t}(s, a) = P_{\mathcal{M_t}, c_t}(s) - P_{\mathcal{M}_t, c_t}(s \setminus a)    
\end{equation}
Where $s \setminus a$ represents the input code sample $s$ without the statement $a$ included. Intuitively, the larger $\Delta P_{\mathcal{M}_t, c_t}(s, a)$ is the more positive impact of the statement towards the prediction of the model $\mathcal{M}_t$.
We employ a greedy strategy and choose the most important statement as the attack location.

\subsubsection{Choosing pattern with meta-model}
Since the number of patterns can be large, the next question is how to choose an effective attack pattern that would likely lead to a different prediction of the model.
For this, we train a decision tree as a meta-model $\mathcal{M}_{meta}: s \mapsto y$. This meta-model is trained to predict the target model prediction $y$, given the presence of each pattern in the pattern set $\mathcal{P}_A$.
In detail, as input to the meta-model, we use a bag-of-pattern encoding.

\noindent \textbf{Obtaining the meta-model}
For each sample source code $s_i$, we construct the feature $\mathbf{f}_i \in \{0, 1\}^{|\mathcal{P}_A|}$. Where:
\begin{equation}
    f_{i, j} = \left\{\begin{array}{ll} 1 & \text{ if } s \text{ contains the pattern $P_j$ }\\ 0 & \text{ otherwise} \end{array} \right.
\end{equation}

Given the features $f_i$ and the corresponding prediction $y_i$ for each source code $s_i$, we can train a decision tree $\mathcal{M}_{meta}$. 
Each path $\pi$ in this decision tree corresponds to a predicted class $c_\pi$ and the number of support $SP_\pi$ (i.e., the number of samples in $\mathcal{D}$ that contains the patterns indicated in the path and received model's prediction to be $c_\pi$).

\noindent \textbf{Choosing pattern}
Given the information on $c_\pi$ and $SP_\pi$ of each path $\pi$, we can now determine which pattern to be inserted into the target source code. 
For each new input $s_{new}$ with AST $T_{new}$, we identify its corresponding features $f_{new}$. 
Furthermore, for each missing pattern $P_{miss}$ in the set of missing patterns $\mathcal{P}_{miss} = \{P_i \in \mathcal{P}_A | P_i \not \subseteq T_{new}\}$, we calculate the approximated probability that adding this missing pattern leads to a different prediction using the meta-model $\mathcal{M}_{meta}$.
\begin{equation}
P\left(\mathcal{M}_{meta}(T_{new} \cup P_{i}) \neq y_i\right) = \frac{\sum_{\pi} SP_{\pi} \times ( \mathbbm{1}_{c_\pi \neq y_i})}{ n }
\end{equation}

Where $\mathbbm{1}_{c_\pi \neq y_i} = \left\{ \begin{array}{ll} 1 & \text{if } c_{\pi} \neq y_i \\ 0 & \text{otherwise} \end{array} \right.$ is the indicator function.
We sample the pattern with the probability of changing model prediction:
\begin{equation}
P_{chosen} = \underset{P_{miss}}{\arg\max}   P\left(\mathcal{M}_{meta}(T_{new} \cup P_{miss}) \neq y_i \right)
\end{equation}

\subsubsection{Pattern insertion}
Having determined the location and pattern to insert, the final question is how to insert the pattern and fill in the $\texttt{<MASK>}$ token such that the final code is syntactically valid and semantically preserving.
To generate syntactically valid code, we leverage a different pre-trained language model $\mathcal{M}_f$ (namely, we use the language-specific CodeBERT provided from CodeBERTScore\cite{zhou2023codebertscore}) to fill in the $\texttt{<MASK>}$. Note that this model is separated from the target model $\mathcal{M}_t$.
In order to make sure the code does not change the semantics of the current code, we follow CARROT's S-modifier \cite{zhang2022Carrot} (which either inserts dead statement or wraps the code inside a redundant branching) and modify the pattern:
\begin{itemize}
    \item If the pattern is conditional (e.g., \texttt{while} loop, \texttt{for} loop, \texttt{if} condition, etc, we modify the original condition of the pattern from \texttt{<MASK>} to \texttt{false \&\& (<MASK>)}).
    \item Else, we put the pattern inside a dead code block.
\end{itemize}
Finally, we insert the modified text pattern into the chosen location and use a pre-trained language model to fill in the mask and obtain the perturbed code. We note that the pre-trained language model might not always generate syntactically valid code. Therefore, we employ tree-sitter parser\footnote{https://github.com/tree-sitter/tree-sitter} to re-parse the generated code and to check if there exist errors in the filled source code. If the tree-sitter detects an erroneous node (i.e., the node has the label ``ERROR''), we retry generating the node up to 5 times then discard the candidate, else, we query the target model to obtain the new evaluation.
The patterns are inserted until either the number of maximum target model queries is met, or the target model changes its prediction.

\section{Experiment Settings}\label{sec:exp_settings}
 The experiment results reported here were obtained on an Intel i5-9600K machine with 64 GB of RAM and equipped with one Nvidia GTX 1080Ti running Linux.

\subsection{Dataset}

In our experiment, we follow the settings of the previous study~\cite{Zhou2022ALERT} and select three downstream tasks from the CodeXGLUE benchmarks~\cite{lu2021codexglue}: Vulnerability Prediction, Clone Detection, and Authorship Attribution. 
Below we introduce the details of each task and its corresponding dataset.

\noindent \textbf{Vulnerability Prediction}.
The objective of this task is to produce a label indicating whether a specified code snippet contains any vulnerabilities. 
Zhou et al.~\cite{zhou2019devign} label source code in two popular open-sourced C projects: FFmpeg\footnote{\url{https://www.ffmpeg.org/}} and Qemu\footnote{\url{https://www.qemu.org/}} to build a dataset consisting of 27,318 functions.
Each function is labeled as either containing vulnerabilities or clean. 
This dataset is widely used to investigate the effectiveness of various code models in understanding code to predict vulnerability. 
We follow the settings in CodeXGLUE to divide the dataset into training, development, and test sets.

\noindent \textbf{Clone Detection}
Clone detection is also modeled as a classification problem: given a pair of two code snippets, a code model should predict whether they are clones (i.e., whether they implement the same function).
We choose BigCloneBench~\cite{Svajlenko2014bigcodebench} as the dataset, which is used in the previous study~\cite{Zhou2022ALERT} and is a widely-acknowledged benchmark for clone detection.
BigCloneBench comprises around 10 million pairs of Java code snippets; over 6 million of them are clones and the remaining 260,000 are not clones.
We create a subset of the dataset that has balanced labels (i.e., the ratio of clone pairs and non-clone pairs is 1:1). 
Following the previous study, we randomly select 90,102 examples for training and 4,000 for validating and testing the code models.

\noindent \textbf{Authorship Attribution}
The task of authorship attribution involves determining the author of a given code snippet. 
We choose the Google Code Jam (GCJ) dataset, which is created using the submission from the Google Code Jam challenge, a yearly global coding competition hosted by Google. 
GCJ dataset is collected and made open-source by Alsulami et al.~\cite{alasumi2017lstmauthorship}, which consists of 700 Python files that are written by 70 authors.
The dataset is balanced, i.e., each author (i.e., class) having 10 code snippets.
This dataset contains mainly Python files but also some C++ code. 
We remove C++ source code to obtain 660 Python files.
We follow an 80:20 split: 20\% of files are used for testing, and 80\% of files are for training.
In accordance with previous studies~\cite{Zhou2022ALERT}, we do not use a validation dataset due to the small dataset size~\cite{Zhou2022ALERT}.

\begin{table}[!t]
    \centering
    \caption{Statistics of tasks and datasets investigated in the paper, as well as the victim models' performance on these datasets. CB and GCB represent CodeBert and GraphCodeBERT, respectively.}
    \label{tab:my_label}
    \setlength{\tabcolsep}{4pt}
    \begin{tabular}{lccccc}
        \toprule
        \textbf{Tasks} & \textbf{Train/Dev/Test} & \textbf{Class} & \textbf{Lang} & \textbf{Model} & \textbf{Acc.} \\
        \midrule
        Vulnerability & \multirow{2}{*}{21,854/2,732/2,732} & \multirow{2}{*}{2} & \multirow{2}{*}{C}  & CB & 63.76\% \\
        Prediction & & &  & GCB & 63.65\% \\ 
        \midrule
        Clone & \multirow{2}{*}{90,102/4,000/4,000} & \multirow{2}{*}{2} & \multirow{2}{*}{Java} & CB & 96.97\% \\ 
        Detection & & & & GCB & 97.36\% \\ 
        \midrule
        Authorship & \multirow{2}{*}{528/-/132} & \multirow{2}{*}{66} & \multirow{2}{*}{Python} & CB & 90.35\%\\ 
        Attribution & & & & GCN & 89.48\%\\
        \bottomrule
    \end{tabular}
\end{table}

\subsection{Target Model, Filler model and Probing Data} 
Following the existing works of ALERT~\cite{Zhou2022ALERT}, we use CodeBERT~\cite{feng2020codebert} and GraphCodeBERT~\cite{guo2021graphcodebert} as our target models. We follow ALERT~\cite{Zhou2022ALERT} in the hyperparameter settings for these models and retrieve the corresponding models from the official GitHub site of ALERT.
For the filler model $\mathcal{M}_f$ which is responsible to fill in the mask, we use language-specific CodeBERTs provided by CodeBERTScore~\cite{zhou2023codebertscore} which has been pre-trained specifically for each language Python, Java and C respectively. For the probing dataset $\mathcal{D}$, we use each task's training source code without the original label.

\subsection{Baselines}
In this study, we compare \tool{} with two state-of-the-art techniques attacking deep code models: CARROT~\cite{zhang2022Carrot} and ALERT~\cite{Zhou2022ALERT}. 
Since CARROT~\cite{zhang2022Carrot} only supports Python, we extend it to attack Python and Java code for sufficient comparison.
Furthermore, CARROT has 4 variants: renaming variables with the model's gradient and by random and inserting dead code guided by the target model's gradient and by random.
Since we follow the black-box settings in our threat model, we use CARROT I-RW (i.e., random identifier renaming) which is the top-performing candidate towards transformer-based models~\cite{zhang2022Carrot}.

\section{Research Questions.} \label{sec:rqs}
To investigate \tool{} against the baselines, we pose three main research questions: (1) The effectiveness and the stealthiness of \tool{} against CARROT and ALERT, (2) Which patterns are the most effective on each task and model, and (3) How adversarial retraining using \tool{} compares with ALERT and CARROT on defending against adversarial attacks. We explain each research question and results below. 

\subsection{RQ1. How effective and stealthy is \tool{} against the state-of-the-art baselines?}
\noindent{\textbf{Effectiveness.}} We use the ALERT's published model as the target model for attacking and compare the Attack Success Rate (ASR) of \tool{} versus CARROT~\cite{zhang2022Carrot} and ALERT~\cite{Zhou2022ALERT}
on the 3 tasks: Authorship Attribution, Vulnerability Prediction, and Clone Detection of CodeXGLUE~\cite{lu2021codexglue}. 
We use the same number of steps and settings to be 2000 following ALERT~\cite{Zhou2022ALERT}. Since \tool{}'s approach needs to execute two code models (the target and the filler) at the same time, it can be slower than the baselines. Thus, in order to give a realistic time efficiency constraint for \tool{}, we also put a timeout of 100 seconds for each attack. We these results below.

\begin{table*}[!t]
\centering
\caption{Attack Success Rate (ASR) comparison for \tool{}, CARROT, and ALERT on CodeXGLUE benchmarks for 3 tasks and 2 models (CodeBERT and GraphCodeBERT). Higher values indicate better performance.}
\begin{tabular}{c|cccccc}
\toprule
\multirow{2}{*}{\textbf{Method}} & \multicolumn{2}{c}{\textbf{Authorship Attribution}} & \multicolumn{2}{c}{\textbf{Vulnerability Prediction}} & \multicolumn{2}{c}{\textbf{Clone Detection}} \\% & \multirow{2}{*}{\textbf{Avg.}} \\
\cmidrule(lr){2-3} \cmidrule(lr){4-5} \cmidrule(lr){6-7}
                & \textbf{CodeBERT} & \textbf{GraphCodeBERT} & \textbf{CodeBERT} & \textbf{GraphCodeBERT} & \textbf{CodeBERT} & \textbf{GraphCodeBERT} \\
\midrule
\tool{}          &    \textbf{0.612}     &  \textbf{0.8407}     &  \textbf{0.774}    &  \textbf{0.799}    &  \textbf{0.401}    &  0.053  \\ % & \textbf{0.579}  \\
CARROT~\cite{zhang2022Carrot}  & 0.485     &   0.598   &  0.620     &  0.746    &  0.108    &   \textbf{0.102}  \\ % \\ & 0.443 \\
ALERT~\cite{Zhou2022ALERT}    &   0.337   &  0.615   &  0.536    &  0.769     & 0.273    &  0.080  \\ % & 0.435 \\
\bottomrule
\end{tabular}
\label{tab:rq1_comparison}
\end{table*}

Table~\ref{tab:rq1_comparison} presents the ASR of \tool{} compared to CARROT and ALERT on the CodeXGLUE benchmarks. 
On average across the three tasks, in terms of attack success rate (ASR), \tool{} outperforms CARROT~\cite{zhang2022Carrot} by $30.6\%$ and ALERT by $33.1\%$ respectively.

In detail, for the target model CodeBERT, \tool{} outperforms CARROT by $26\%$, $24\%$ and ALERT by $81.5\%$, $43.6\%$ on Authorship Attribution and Vulnerability Prediction respectively. 
For GraphCodeBERT, the corresponding improvements are $40.5\%$ and $7\%$ for CARROT and $36.65\%$ and $3.8\%$ for ALERT.
For Clone Detection and on CodeBERT, \tool{} outperformed ALERT by $46.9\%$ and CARROT by $270\%$ respectively. On GraphCodeBERT, \tool{} performs comparably with CARROT and ALERT. This better performance can be attributed to \tool{}'s capability in leveraging specific attack patterns towards the target model. Pre-trained language models of code rely on both syntactic patterns and textual tokens \cite{karmakar2021pretrained}. The key difference between \tool{}, ALERT, and CARROT is that \tool{} adds varied code fragments, while ALERT and CARROT I-RW only change existing variable names. Since variables are only part of the model input, \tool{}'s adding new varying code fragments expands the attack space and results in more comprehensive model perturbation.

The performance of all attack methods changes when switching from CodeBERT to GraphCodeBERT: On GraphCodeBERT~\cite{guo2021graphcodebert}, \tool{} marginally outperforms ALERT~\cite{Zhou2022ALERT} and CARROT~\cite{zhang2022Carrot}, which can be attributed to two factors: (1) GraphCodeBERT's emphasis on variable names, and (2) the perturbations caused by the three tools. Recall that GraphCodeBERT augments CodeBERT with explicit variable names and attention masks, making CARROT and ALERT's renaming more effective. At the same time, since GraphCodeBERT's dataflow does not filter out dead branches, \tool{} attacks leveraging surrounding variables can still alter the dataflow graph and slightly outperform the baselines.

While the ASR on CodeBERT clone detection by \tool{} is nearly 42\% better than ALERT and four times better than CARROT (i.e., success rates of 0.4, 0.27, 0.1 respectively). The success rates on GraphCodeBERT are low. This suggests that there is still room for improvements on the code clone detection task. 

\noindent{\textbf{Stealthiness.}} 
"Stealthiness" measure how hard it is for the developer to notice the attack. Intuitively, the closer the resulting attacked source code is to the original source code, the harder it is to notice, hence, the attack is stealthier. As a proxy to measure stealthiness automatically, we use \textit{Code Change Rate}. The lower the code change rate is, the stealthier the attacks. Assume that each source codes $s$ is tokenized into $n_t$ number of tokens, and the attack required $n_i$ tokens to be modified. We calculate the average and standard deviation $\mu_{TC}$ and $\sigma_{TC}$ of the number of inserted tokens as well as the change rate of $n_i/n_t$.

The change rate of \tool{} is shown in Table~\ref{tab:rq2_comparison}. 
On all 3 tasks, \tool{} needs to insert approximately 100 tokens. For CodeBERT model, \tool{} inserts on average 90.27 tokens with a standard deviation of 58.38 for Authorship Attribution, 56.33 (38.96) for Vulnerability Prediction, and 125.29 (72.98) for Clone Detection. The corresponding change rates are 0.136, 0.570, and 0.1455.
On GraphCodeBERT, \tool{} has average change rates of 112.833, 62.06, and 124 inserted tokens with standard deviations of 64.4, 47.165, and 70.2 for the three tasks respectively. The corresponding change rates are 0.1359, 0.463, and 0.144.
\tool{}'s number of average inserted tokens is smaller in comparison with the number of changed tokens from ALERT and CARROT, as well as having lower variation in the number of inserted tokens across all tasks.  Moreover,  \tool{}'s change rate is comparable to ALERT and CARROT on Authorship attribution and Clone Detection but is higher in Vulnerability Detection. This is due to identifier renaming methods' change rates grow with the number of variable usages in the code. Since Vulnerability Prediction's source code is smaller than the two other tasks, this per-source code change rate is higher, on the contrary, when the program size grows, \tool{} demonstrates better change rates.

\begin{table*}[ht]
\centering
\caption{Code Change Rate of \tool{}, ALERT, and Carrot, $\mu_{TC}$ and $\sigma_{TC}$ are the total number of tokens added and the standard deviation. $\mu_{TCR}$ and $\sigma_{TCR}$ are the token change rate and the standard deviation respectively}
\begin{tabular}{cccccccccccccc}
\toprule
Method & Target Model & \multicolumn{4}{c}{Authorship Attribution} & \multicolumn{4}{c}{Vulnerability Prediction} & \multicolumn{4}{c}{Clone Detection} \\ 
\cmidrule(lr){3-6} \cmidrule(lr){7-10} \cmidrule(lr){11-14}
 & & \(\mu_{TC}\) & \(\sigma_{TC}\) & \(\mu_{TCR}\) & \(\sigma_{TCR}\) & \(\mu_{TC}\) & \(\sigma_{TC}\) & \(\mu_{TCR}\) & \(\sigma_{TCR}\) & \(\mu_{TC}\) & \(\sigma_{TC}\) & \(\mu_{TCR}\) & \(\sigma_{TCR}\) \\ 
\midrule
GCA & CB & 90.27 & 58.38 & 0.136 & 0.149 & 56.33 & 38.96 & 0.570 & 0.378 & 98.13 & 97.98 & 0.255 & 0.367 \\ 
 & GCB & 112.833 & 64.4 & 0.1359 & 0.1724 & 62.06 & 47.165 & 0.463 & 0.794 & 40.75 & 33.75 & 0.03 & 0.04 \\ 
\midrule
ALERT & CB & 151.95 & 144.254 & 0.1416 & 0.102 & 136.14 & 320.18 & 0.1295 & 0.086 & 69.68 & 97.191 & 0.112 & 0.0644 \\ 
 & GCB & 325.94 & 195.98 & 0.344 & 0.124 & 159.302 & 371.39 & 0.121 & 0.086 & 153.36 & 193.34 & 0.264 & 0.1376 \\ 
\midrule
CARROT & CB & 113.65 & 120.62 & 0.24 & 0.193 & 107.133 & 209.8 & 0.136 & 0.149 & 88.87 & 232 & 0.103 & 0.121 \\ 
 & GCB & 129.101 & 161.29 & 0.26 & 0.18 & 112.61 & 179.61 & 0.215 & 0.2 & 170.95 & 297.866 & 0.2344 & 0.168 \\ 
\bottomrule
\end{tabular}
\label{tab:rq2_comparison}
\end{table*}

\begin{graybox}
\textbf{RQ1 Conclusion:} \tool{} outperforms CARROT and ALERT in Authorship Attribution and Vulnerability Prediction tasks, with similar results in Clone Detection. For stealthiness, \tool{} achieves reasonable change rates. \tool{} gives a better code change rate on larger source codes while ALERT and CARROT's change rates grow with the length of the input code.
\end{graybox}

\subsection{RQ2. What are the most effective patterns on each problem and model?}

To understand which patterns contribute the most to the success of adversarial attacks, we evaluate the frequency with which a pattern is added in successful adversarial examples. We report the Top-3 most frequently occurring patterns for task and model combination.
We do not count the dead code wrapper (e.g., if False) since they are not the original patterns.
Since the tasks are done in different languages, we group the equivalent AST between different languages to count the patterns' frequency. For example, \texttt{comparison\_operator} in Python between two expressions is equivalent to a \texttt{binary\_expression} in C++ and Java, \texttt{block} in Python is equivalent to \texttt{compound\_statement} in C++ and Java. If an AST pattern only appears in a single language, we report the original pattern.

% If there is partial of the complete AST, we group the into the super pattern.

% 80% of them contain the if statement
% GCB: 80% related to function call
% GCB is robust to the changes in the code structure but not robust to variable renaming.

\begin{figure}
\centering
\includegraphics[width=\linewidth]{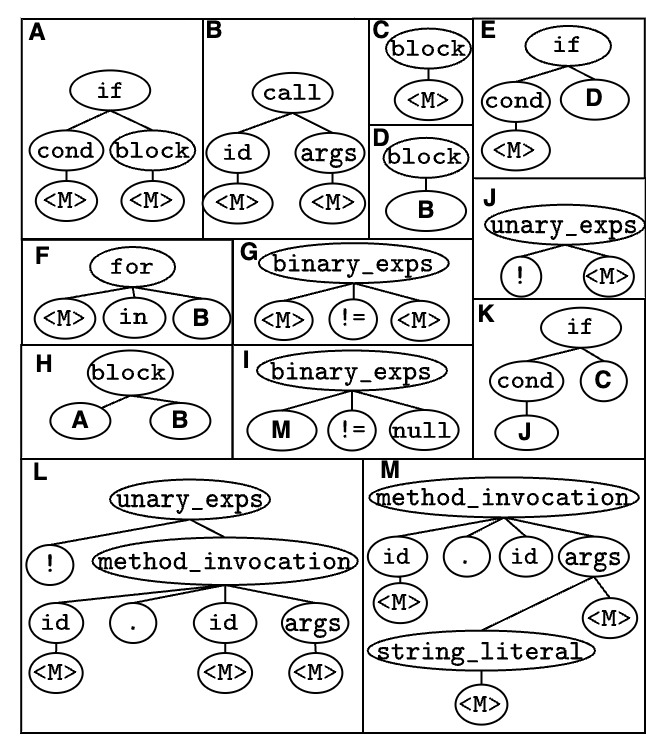}
\caption{Top frequent patterns among attacks}
\label{fig:patterns}
\end{figure}

\begin{table}[htbp]
\centering
\caption{Top frequent patterns in a successful attack, CF means the patterns contain a control-flow element (e.g., \texttt{if, else-if, else, for, while, etc.}) and DF means the patterns contain calculations relating to variables (e.g., identifier nodes), L means the patterns contains Literals}
\begin{tabular}{@{}cc|c|c|c@{}}
\toprule
\textbf{Task} & \textbf{Model} & \textbf{Pattern ID} & \textbf{Pattern type} & \textbf{Frequency} \\
\midrule
 & \multirow{3}{*}{CB} & A & CF & 0.987 \\
& & G & DF & 0.703 \\
\textbf{Authorship} & & B & CF, DF & 0.604 \\ 
\cmidrule{2-5} 
\textbf{Attribution} & \multirow{3}{*}{GCB} & B & CF, DF & 0.586 \\
 & & D & DF & 0.432 \\
 & & F & CF, DF & 0.211 \\
 \midrule
& \multirow{3}{*}{CB} & E & CF & 0.864\\ 
& & K & CF &  0.807 \\
\textbf{Vulnerability} & & I & CF & 0.174 \\\cmidrule{2-5} 
 \textbf{Prediction} & \multirow{3}{*}{GCB} & B & CF, DF & 0.413\\ 
 & & H & CF, DF & 0.283  \\
& & E & CF, DF & 0.249 \\
\midrule 
& \multirow{3}{*}{CB} & G & DF & 0.448 \\
 & & L & CF, DF & 0.085 \\
\textbf{Clone} & & M & DF, L & 0.023 \\ \cmidrule{2-5}
\textbf{Detection} & \multirow{3}{*}{GCB} & G & DF & 0.615\\
 & & K & CF, DF & 0.482 \\
& & E & CF, DF & 0.448 \\
\bottomrule
\end{tabular}
\end{table}

Table~\ref{tab:rq2_comparison} and Figure~\ref{fig:patterns} present the top frequently appearing patterns in successful attacks for each task and target model and Figure~\ref{fig:patterns} depicts these patterns in detail. The results indicate that patterns' effectiveness varies across different tasks and target models. 
Particularly on all 3 tasks, CodeBERT has the 2 top frequently appeared patterns containing control-flow elements and 1 top pattern that contains data-flow elements.
While GraphCodeBERT also has $2/3$ top patterns containing control-flow elements, all of its top-frequented patterns in the successful attack contain data-flow elements. This hints that GraphCodeBERT models rely more on variable and data-flow operations (which is consistent with the design of the model).

Among different tasks, Authorship Attribution and Clone Detection both have $4/6$ patterns containing control flow and $5/6$ patterns using data flow elements.
Vulnerability Prediction has $6/6$ patterns containing control flow elements and $3/6$ containing data flow elements.
This suggests that Authorship Attribution and Clone Detection have more reliance on data flow elements while the Vulnerability Prediction model leans more toward the control flow elements.

\begin{graybox}
\textbf{RQ2 Conclusion:} 
Among the most frequently occurring patterns, CodeBERT frequently exhibits control-flow patterns, while GraphCodeBERT relies more on data-flow operations. Authorship Attribution and Clone Detection depend on data flow elements, whereas Vulnerability Prediction leans towards control flow elements.
\end{graybox}

\subsection{RQ3. How effective is \tool{} in improving the robustness of code models?}

%Since obtaining all successful attack examples according to ALERT~\cite{Zhou2022ALERT} and CARROT~\cite{zhang2022Carrot} can be time-consuming as we have to query both the original model $\mathcal{M}_t$ and the filler model $\mathcal{M}_f$ multiple times, we instead, 

\noindent \textbf{Settings.} To answer this question, we experiment with adversarial retraining, a process that involves fine-tuning the target model with adversarial examples to improve its robustness against potential attacks. 
To do this, we randomly insert the patterns generated by \tool{} into the training dataset as a data augmentation step. 
While training, on each data sample, we set the probability of applying a perturbation to $0.5$. This probability indicates how likely we would apply a perturbation on each sample. Setting this probability to $0.5$ balances the use of original and perturbed samples in the training process.
Recall that \tool{} can insert dead code multiple times, we also set the maximum number of perturbations to $5$, meaning that we will insert the dead code for the maximum of $5$ times, which is half the number of greedy steps that \tool{} uses in attacking.

We obtain the ALERT~\cite{Zhou2022ALERT}'s fine-tuned model from the official GitHub repository\footnote{\url{https://github.com/soarsmu/attack-pretrain-models-of-code}}. 
For CARROT~\cite{zhang2022Carrot}, we follow the original procedure and obtain the perturbed samples that either change the model's prediction or drop the target model's confidence towards the original prediction, we augment the training set with these samples and fine-tune until the model converges.

We test the robustness of the fine-tuned models on the adversarial samples in RQ1 of both ALERT, CARROT, and \tool{} itself.
The robustness measurement of a single adversarial sample is defined according to CARROT~\cite{zhang2022Carrot}: we measure the ratio of making correct predictions on the set of generated adversarial examples in RQ1.

\begin{table*}[htbp]
\centering
\caption{Robustness improvement of the target model after adversarial fine-tuning: Each major column indicates from which method the adversarial examples are generated, and each minor column indicates from which method the model is adversarially fine-tuned.}
\begin{tabular}{@{}ll|ccc|ccc|ccc@{}}
\toprule
\textbf{Task} & \textbf{Model} & \multicolumn{3}{c|}{\textbf{CARROT}} & \multicolumn{3}{c|}{\textbf{ALERT}} & \multicolumn{3}{c}{\textbf{\tool{}}} \\ 
\cmidrule(lr){3-5} \cmidrule(lr){6-8} \cmidrule(lr){9-11}
& & CARROT & ALERT & GCA & CARROT & ALERT & GCA & CARROT & ALERT & GCA \\ 
\midrule
Authorship & CodeBERT & 0.2528 & 0.6781 & 0.3793 & 0.8095 & 0.8736 & 0.8095 & 0.2916 & 0.6067 & 0.7661\\ 
Attribution & GraphCodeBERT & 0.00 & 0.0253 & 0.0253 & 0.5143 & 0.9621 & 0.9143 & 0.0096 & 0.027 & 0.4144\\
\midrule
Vulnerability & CodeBERT & 0.5145 & 0.5364 & 0.495 & 0.8518 & 0.8811 & 0.7362 & 0.5957 & 0.5786 & 0.6047 \\ 
Prediction & GraphCodeBERT & 0.5635 & 0.5242 & 0.5257 & 0.7966 & 0.8904 & 0.7893 & 0.578 & 0.582 & 0.5995\\
\midrule
Clone & CodeBERT & 0.9568 & 0.9606 & 0.9722 & 0.9012 & 0.9190 & 0.9120 & 0.9232  & 0.9341 & 0.9674\\
Detection & GraphCodeBERT & 0.9434 & 0.9032 & 0.9322 & 0.9123 & 0.9104 & 0.9089 & 0.9233 & 0.9142 & 0.9258 \\
\bottomrule
\end{tabular}
\label{tab:rq3_robustness}
\end{table*}

\noindent \textbf{Results.}
The results of adversarial training are presented in Table~\ref{tab:rq3_robustness}.  
Overall, on Authorship Attribution, using the Wilcoxon Rank Sum Test, we obtained a $p$-value of 0.031 for CARROT vs ALERT, 0.043 for CARROT vs \tool{}, and 0.893 between ALERT and \tool{}. The Cliff's Delta for the three pairs (CARROT vs ALERT, CARROT vs \tool{} and ALERT vs \tool{}) are $-0.444$ (medium), $-0.472$ (medium), and $0.028$ (small) respectively. This demonstrates that ALERT and \tool{} exhibit similar performance in improving robustness, and both methods outperform CARROT. 
It is worth noting that for both Vulnerability Prediction and Clone Detection, the robustness improvements across different adversarial fine-tuning methods are similar for all baselines.

Interestingly, in the case of Authorship Attribution and Vulnerability Prediction, both CARROT and \tool{}'s adversarial results are difficult to defend against. For CARROT, this can be attributed to the fact that changes in variables can be more pronounced than in ALERT, owing to the lack of natural constraints. For \tool{}, the difficulty in defense may be due to the variability of \tool{}'s attacks: since the inserted attacks in \tool{} are filled using a pre-trained language model, the filled masks may vary, making it challenging to defend against the produced attacks.

Finally, while retraining with \tool{} and ALERT results in a similar defense, we emphasize that \tool{} only employs an augmentation method instead of performing a full adversarial attack on the target model or the target model's response itself, giving it finetuning an advantage of efficiency and in deployment.
% For GraphCodeBERT, when fine-tuned with \tool{}, the model's robustness against ALERT and \tool{} adversarial samples in Authorship Attribution task is 0.9143 and 0.4144, respectively. When fine-tuned with ALERT, the robustness scores are 0.9621 and 0.027. In this case, \tool{} fine-tuning significantly improves the model's resilience against \tool{} attacks but reduces its robustness against ALERT attacks. ALERT fine-tuning, on the other hand, provides better robustness against ALERT attacks but fails to protect against \tool{} attacks.

\begin{graybox}
\textbf{RQ3 Conclusion:} 
\tool{}'s fine-tuning, although not requiring the target models' feedback nor a full adversarial attack on the training dataset like ALERT and CARROT, achieves similar performance with ALERT and better performance in comparison with CARROT.
\end{graybox}

\section{Threats to Validity}

\subsection{Threats to Internal Validity}
The effectiveness of \tool{} relies on the mined AST patterns. If the pattern mining process is not comprehensive or biased towards specific patterns, it may affect the success rate of our adversarial attacks. 
 We mitigate this threat by using a model-agnostic approach to mine discriminative patterns and validating our method against multiple datasets and model architectures.
The process of inserting the AST patterns into the code and the selection of target models may introduce randomness, which can potentially influence the results. 
We address this issue by conducting experiments with 3 runs and reporting the average performance, ensuring the stability and reliability of our results. 

\subsection{Threats to External Validity}
The generalizability of our findings might be limited by the choice of datasets, models, and evaluation metrics used in our experiments. To mitigate this threat, we employed widely-used tasks and datasets that are included in the CodeXGLUE benchmark, as well as the popular pre-trained language models for code: CodeBERT and GraphCodeBERT. 

\subsection{Threats to Construct Validity}
The choice of evaluation metrics can impact the interpretation of our results. In this study, we used the Attack Success Rate (ASR) to measure the effectiveness of our method, which was widely used in the previous state-of-the-art papers~\cite{Zhou2022ALERT, zhang2022Carrot}.
Moreover, we also adopt the same metric used to assess the robustness improvement from previous studies~\cite{Zhou2022ALERT, zhang2022Carrot}.

\section{Related Work}
\label{sec:rel_work}

This section provides an overview of the relevant studies in the field. 
We divide these related works into two categories: (1) pre-trained models of code and (2) potential threats to these models.

\subsection{Pre-trained Models of Code}

Large language models, such as the BERT~\cite{bert,RoBERTa} and GPT~\cite{gpt-2,gpt-3} families, have achieved remarkable performance in various natural language processing tasks. This success inspired researchers to develop pre-trained models for programming languages to capture code semantics and improve code-related tasks.

The trend began with CodeBERT~\cite{feng2020codebert}, based on RoBERTa~\cite{RoBERTa}, which was pre-trained on the bimodal CodeSearchNet dataset~\cite{husain2019codesearchnet}. It has two pre-training objectives: masked language modeling and replaced token detection. GraphCodeBERT~\cite{guo2021graphcodebert} further incorporates code graph structure, adding data flow edge prediction and node alignment. Other models, such as CuBERT~\cite{CuBERT} and C-BERT~\cite{CBERT}, focus on Python and C source code, respectively. These \textit{encoder-only} models generate code embeddings for downstream tasks.

Another type of code model is \textit{decoder-only} models, primarily focused on code generation tasks. The GPT-based code model is a well-known decoder-only architecture. Lu et al. introduce CodeGPT in the CodeXGLUE benchmark~\cite{lu2021codexglue}, utilizing the GPT-2 architecture and pre-trained on CodeSearchNet~\cite{husain2019codesearchnet}. Larger models include InCoder~\cite{fried2023incoder} and CodeGen~\cite{codegen} with 16.1B parameters. OpenAI's Codex~\cite{codex} powers Microsoft CoPilot. A recent study suggests that smaller models trained on high-quality datasets can outperform larger models~\cite{santacoder}.

Researchers have also applied \textit{encoder-decoder} architecture to code models. Inspired by BART~\cite{bart} and T5~\cite{T5}, they propose models like PLBART~\cite{plbart} and CodeT5~\cite{wang2021codet5}, experimenting with pre-training tasks such as masked span prediction and masked identifier prediction. Other models include DeepDebug~\cite{DeepDebug}, Prophetnet-x~\cite{prophetnet}, CoTexT~\cite{cotext}, and SPT-Code~\cite{spt-code}. These code models demonstrate remarkable performance on various code-related tasks, including code completion, code summarization, and code generation~\cite{niu2023empirical}. 
% For an empirical comparison of these pre-trained code models, refer to~\cite{niu2023empirical}.
% {\color{cyan}After deciding our victim models, we need to explain why choose them here briefly.}

\subsection{Threats to Code Models}
Despite their impressive performance, code models remain vulnerable to various attacks. Understanding these vulnerabilities is crucial for enhancing their security and protecting their downstream applications.

One significant threat is the susceptibility of code models to adversarial examples~\cite{pmlr-v119-bielik20a}.
Yefet et al.~\cite{Yefet2020} were among the first to study adversarial robustness in code models, employing FGSM~\cite{FGSM} to rename variables and target code models like code2vec~\cite{code2vec}.
Jordan et al.~\cite{9825895} extend \cite{Yefet2020} work by considering additional transformations such as converting \texttt{for} loop to \texttt{while} loop. 
Srikant et al.~\cite{Epresentation2021} utilize PGD~\cite{PGD} to generate stronger adversarial examples.
These studies assume \textit{white-box} access to the code models, meaning that the attacker has access to the model's parameters and gradient information.

Attacks can also be conducted in a \textit{black-box} manner. 
Zhang et al.~\cite{MHM} propose Metropolis-Hastings Modifier (MHM) to for black-box adversarial example generation.
Rabin et al.~\cite{rabin2021generalizability} and Applis et al.~\cite{9678706} use semantic-preserving and metamorphic transformations, respectively, to assess code model robustness.
Tian et al.~\cite{9724884} employ reinforcement learning for attacks, while Jia et al.~\cite{jia2023clawsat} demonstrate that adversarial training improves code model robustness and correctness.
Pour et al.~\cite{pour2021searchbased} proposed leveraging renaming variables, argument, method, and API names as well as adding argument, print statement, for loop, if loop, and changing returned variables to generate adversarial source codes to improve code models' robustness.
Yang et al.~\cite{Zhou2022ALERT} emphasize the naturalness requirement for code model adversarial examples and develop ALERT, which uses genetic algorithms for example generation. 
Zhang et al.~\cite{zhang2022Carrot} propose CARROT, which employs worst-case performance approximation to measure code model robustness. Both ALERT and CARROT demonstrate state-of-the-art performance. \tool{} focuses on crafting more complex adversarial examples rather than emphasizing naturalness.

Researchers have also studied threats like data poisoning and backdoor attacks. Ramakrishnan et al.~\cite{codebackdoor} propose fixed and grammar triggers to insert backdoors into code models, while Wan et al.~\cite{you-see} inject similar triggers into code search models. Yang et al.~\cite{advdoor} use adversarial examples for stealthy backdoor injection, and Li et al.~\cite{li2022poison} generate dynamic triggers using language models, proposing an effective defense method. Schuster et al.~\cite{263874} conduct data poisoning for insecure API usage, while Nguyen et al.~\cite{coffee} assess the risk of malicious code injection in API recommender systems. Sun et al.~\cite{CoProtector} demonstrate data poisoning can protect open-source code from unauthorized training.

\section{Conclusion and Future Work}
We presented \tool{}, an adversarial attack tool for pre-trained code models to better evaluate the robustness of code models. 
We evaluate the robustness of two popular code models (e.g., CodeBERT and GraphCodeBERT) against our proposed approach on three tasks: Authorship Attribution, Vulnerability Prediction, and Clone Detection. The experimental results suggest that our proposed approach significantly outperforms state-of-the-art approaches in attacking code models such as CARROT and ALERT. Based on the average attack success rate (ASR), \tool{} achieved $30\%$ improvement over CARROT and $33\%$ improvement over ALERT respectively. 
We also evaluate the produced attack quality with the usage of code change rate and shows that \tool{} produces successful attacks with fewer token change in general and the code change rate decrease with larger code files.
Furthermore, \tool{}'s adversarial fine-tuning has a similar performance with ALERT in enhancing model robustness, while requiring neither the target model's output nor conducting a full adversarial attack on the training data.

For future work, we plan to investigate the impact of different perturbations to improve \tool{}'s performance. Additionally, we aim to explore more attack scenarios, such as multi-label and multi-class settings, to further evaluate the effectiveness of \tool{} and enhance its generalizability across a wider range of code-related tasks.

% \section{Reproduction package}

% \begin{tcolorbox}[colback=white, colframe=black, boxrule=0.4pt]
%     In the interest of responsible disclosure, we must emphasize that the open-source code and documentation
%     at \textbf{\url{https://doi.org/10.6084/m9.figshare.22774781}} are provided to support reproducibility.
% However, they should not be misused for harmful activities, including attacking deployed code models.
% \end{tcolorbox}

\bibliographystyle{ACM-Reference-Format}

\onecolumn
\begin{multicols}{2}
\bibliography{library}
\end{multicols}

\end{document}